\documentclass[conference,9pt]{IEEEtran}

\usepackage{algpseudocode}
\algtext*{EndWhile}
\algtext*{EndIf}
\algtext*{EndFor}
\algtext*{EndProcedure}
\usepackage{amsmath}
\usepackage{amssymb}
\usepackage{array}
\usepackage{algorithm}
\usepackage{bm}
\usepackage[bookmarks=false, hidelinks]{hyperref}
\usepackage[noadjust]{cite}
\usepackage[font=small]{caption}
\usepackage{color}
\usepackage{comment}
\usepackage{dblfloatfix}
\usepackage[inline]{enumitem}
\usepackage{epsfig}
\usepackage{etoolbox}
\usepackage{fancybox}
\usepackage{float}
\usepackage{graphicx}
\usepackage{listings}
\usepackage{mathrsfs}
\usepackage{multirow}
\usepackage{multicol}
\usepackage{pifont}
\usepackage{placeins}
\usepackage{rotating}
\usepackage{setspace}
\usepackage{soul}
\usepackage[hang, tight]{subfigure}
\usepackage{threeparttable}
\usepackage{times}
\usepackage{url}
\usepackage{upgreek}
\usepackage{verbatim}
\usepackage{wrapfig}
\usepackage[usenames,dvipsnames]{xcolor}
\usepackage{authblk}
\usepackage[]{siunitx}
\usepackage[super]{nth}
\usepackage{tabularx}
\usepackage{supertabular}
\usepackage{tikz}
\usepackage{bbm}
\usepackage{blindtext}
\usepackage{booktabs}
\usepackage{diagbox}
\usepackage{tabulary}
\usepackage{tablefootnote}
\usepackage[dvipsnames]{xcolor}

\usepackage{tikz}
\usepackage{xcolor}

\usepackage{listings}
\usepackage{xcolor}
\usepackage[hidelinks]{hyperref}

\definecolor{codegreen}{rgb}{0,0.6,0}
\definecolor{codegray}{rgb}{0.5,0.5,0.5}
\definecolor{codepurple}{rgb}{0.58,0,0.82}
\definecolor{backcolour}{rgb}{0.95,0.95,0.92}

\lstdefinestyle{mystyle}{
  backgroundcolor=\color{backcolour},   commentstyle=\color{codegreen},
  keywordstyle=\color{magenta},
  numberstyle=\tiny\color{codegray},
  stringstyle=\color{codepurple},
  basicstyle=\ttfamily\footnotesize,
  breakatwhitespace=false,         
  breaklines=true,                 
  captionpos=b,                    
  keepspaces=true,                 
  numbers=left,                    
  numbersep=5pt,                  
  showspaces=false,                
  showstringspaces=false,
  showtabs=false,                  
  tabsize=2
}
\usepackage{colortbl}
\usepackage{dcolumn}

\definecolor{greenDeep}{RGB}{0,170,0}
\definecolor{greenSlightDeep}{RGB}{0,205,0}
\definecolor{greenShallow}{RGB}{0,255,0}
\definecolor{greenShallower}{RGB}{160,255,0}
\definecolor{orangeShallow}{RGB}{255,190,0}
\definecolor{orangeDeep}{RGB}{255,80,0}
\definecolor{orangeDeeper}{RGB}{255,40,0}
\definecolor{redDeep}{RGB}{255,0,0}

\definecolor{redLight}{RGB}{255,128,114}

\def\zz#1{%
\ifdim#1pt>4.9pt\cellcolor{greenDeep}\else
\ifdim#1pt>3.9pt\cellcolor{greenSlightDeep}\else
\ifdim#1pt>2.9pt\cellcolor{greenShallower}\else
\ifdim#1pt>2.9pt\cellcolor{yellow}\else
\ifdim#1pt>1.9pt\cellcolor{orangeShallow}\else
\ifdim#1pt>1.9pt\cellcolor{orange}\else
\ifdim#1pt>0.9pt\cellcolor{orange}\else
\ifdim#1pt>0.9pt\cellcolor{orangeDeep}\else
\cellcolor{orangeDeep}\fi\fi\fi\fi\fi\fi\fi\fi
#1}

\graphicspath{{./fig/}}

\makeatletter
\renewcommand\footnoterule{%
  \kern-3\p@
  \hrule\@width0.4\columnwidth
  \kern2.6\p@}
  \makeatother

\newcommand{\question}[1]{\textcolor{orange}{[#1]}}

\usepackage[hang,flushmargin]{footmisc}

\begin{document}


\title{ATLAS: \underline{A} Self-Supervised and Cross-S\underline{t}age Net\underline{l}ist Power Model for Fine-Grained Time-Based L\underline{a}yout Power Analysi\underline{s}\vspace{-.1in}}

\author{ 
Wenkai Li$^\dagger$, Yao Lu$^\dagger$, Wenji Fang, Jing Wang, Qijun Zhang, Zhiyao Xie$^*$\\ 
Hong Kong University of Science and Technology\\
\{wlidm, yludf, wfang838, jwangjw, qzhangcs\}@connect.ust.hk, eezhiyao@ust.hk 
\\\fontsize{10}{10}\selectfont ($^\dagger$equal contribution, $^*$corresponding author)}

\maketitle

\begin{abstract}

Accurate power prediction in VLSI design is crucial for effective power optimization, especially as designs get transformed from gate-level netlist to layout stages. However, traditional accurate power simulation requires time-consuming back-end processing and simulation steps, which significantly impede design optimization. To address this, we propose ATLAS, which can predict the ultimate time-based layout power for any new design in the gate-level netlist.  
To the best of our knowledge, ATLAS is the first work that supports both time-based power simulation and general cross-design power modeling. It achieves such general time-based power modeling by proposing a new pre-training and fine-tuning paradigm customized for circuit power.
Targeting golden per-cycle layout power from commercial tools, our ATLAS achieves the mean absolute percentage error (MAPE) of only 0.58\%, 0.45\%, and 5.12\% for the clock tree, register, and combinational power groups, respectively, without any layout information. Overall, the MAPE for the total power of the entire design is $<$1\%, and the inference speed of a workload is significantly faster than the standard flow of commercial tools. 

\end{abstract}

\section{Introduction}\label{sec:intro}

\begin{figure}[!t]
\centering
\includegraphics[width=0.46\textwidth]{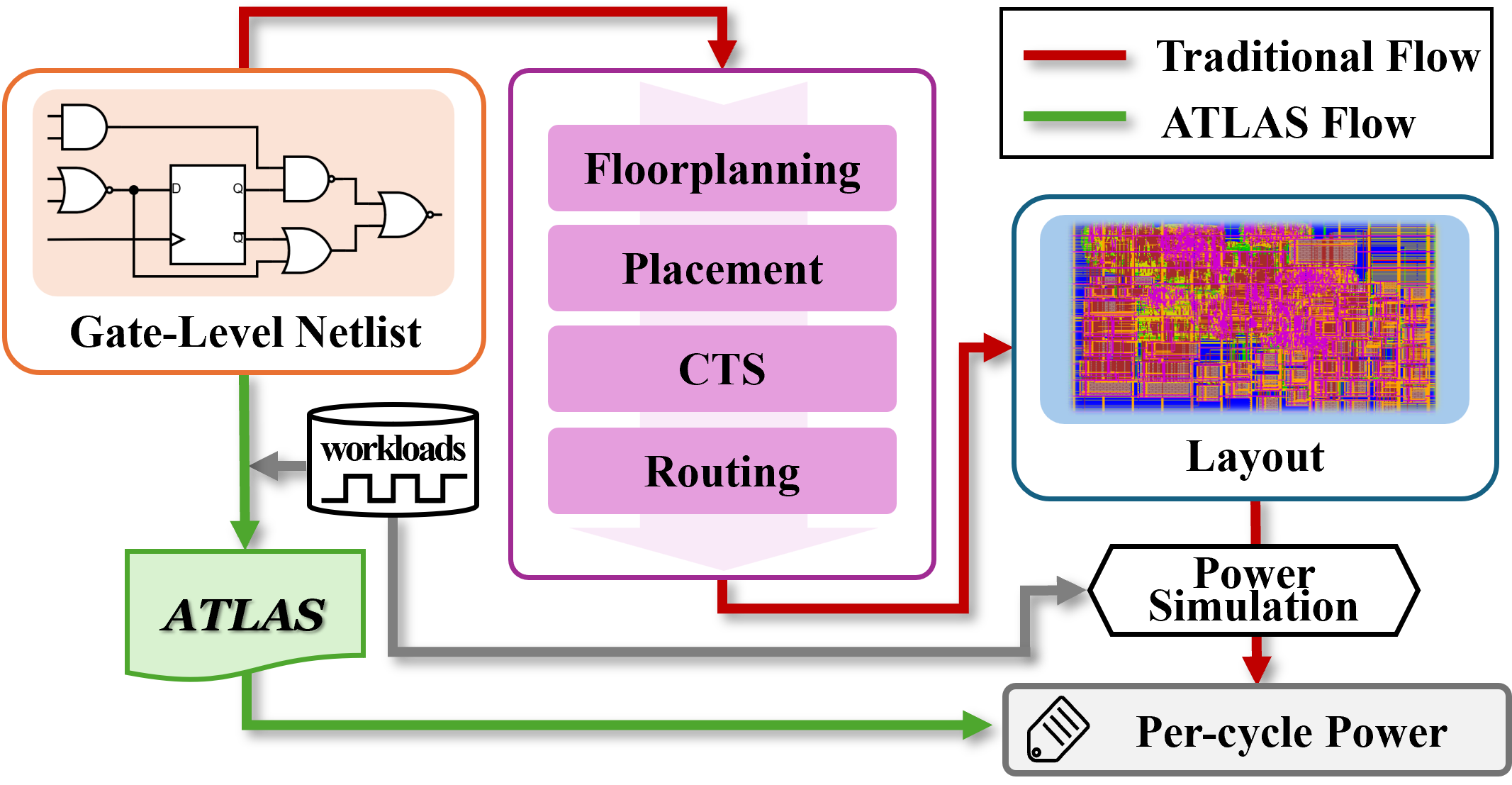}
\caption{Overview of ATLAS for time-based netlist power modeling. Standard traditional power simulation for post-layout design is time-consuming due to both layout steps and per-cycle power simulation. ATLAS achieves significant acceleration over the standard flow.\vspace{-.1in}}
\vspace{-.1in}
\label{fig:flow}
\end{figure}

Power is an increasingly important objective in modern chip design. As design complexity keeps scaling up, it is increasingly costly for chip designers to get accurate power values of their design. It is even more challenging to simulate time-based (e.g., per-cycle) power values, which enables the analysis of peak power and power fluctuations ($Ldi/dt$).  
As Fig.~\ref{fig:flow} shows, starting at a gate-level netlist, designers need to complete the design whole layout process and then employ target workloads (in .fsdb or .vcd format) to simulate time-based power consumption based on commercial EDA tools~\cite{ptpx, powerpro}. Both layout and power simulation can take days or even weeks for large designs or workloads. In summary, accurate, efficient, time-based power models are in high demand.


\textbf{Prior Power Modeling Works.} In recent years, various novel data-driven power modeling techniques~\cite{zhou2019primal, xie2021apollo, sengupta2022good, xu2022sns, xu2023fast, fang2023masterrtl, du2024powpredi} have been proposed, providing unprecedented early-stage and fast power evaluations based on machine learning (ML) power models. Representative power modeling works have been summarized in Table~\ref{tbl:prior work}. However, no prior works can provide \emph{time-based power} and generalize across \emph{different designs} simultaneously. We categorize existing power modeling solutions into two main types: 
\begin{itemize}
    \item The first type of works (e.g., PRIMAL~\cite{zhou2019primal}, APOLLO~\cite{xie2021apollo}) can provide accurate and time-based power values, but they do not support general cross-design models. Instead, they~\cite{zhou2019primal, xie2021apollo} require training a new design-specific power model from scratch for each new design. This development process, especially the label collection step, is highly time-consuming. 
    \item The second type of works~\cite{sengupta2022good, xu2022sns, xu2023fast, fang2023masterrtl, du2024powpredi} can provide a general power model that applies to new designs, which are unknown during model training. Perhaps due to the difficulty of generalization, none of these works further provide time-based power values. Moreover, most of these works~\cite{sengupta2022good, xu2022sns, xu2023fast, du2024powpredi} do not model power based on different workloads. Instead, they only model the average power value based on the propagation of user-defined toggle rates (i.e., vectorless power values).  
\end{itemize}

In addition, many prior power modeling solutions~\cite{zhou2019primal, xie2021apollo, sengupta2022good, xu2022sns, xu2023fast} do not model the ultimate power after design layout. For faster data collection, they adopt power values simulated at the gate-level netlist stage as labels, skipping the layout process. As a result, they are not validated on capturing ultimate post-layout power, which is heavily affected by many factors such as accurate metal wire capacitance, timing optimization on netlist (e.g., buffer insertion, netlist reconstruction), the clock tree, etc.

\begin{table}[!b]
\vspace{-.10in}
\centering
 \renewcommand{\arraystretch}{1.2}
 \resizebox{.5\textwidth}{!}{
\begin{tabular}{|c|c|c|c|c|c|}
\hline 
Power Models &
  \begin{tabular}[c]{@{}c@{}}Applied \\ Stage\end{tabular} &
  \begin{tabular}[c]{@{}c@{}}Support\\ Workloads\end{tabular} &
  \begin{tabular}[c]{@{}c@{}}Time- \\ Based\end{tabular} &
  \begin{tabular}[c]{@{}c@{}}Cross- \\ Design\end{tabular} &
  \begin{tabular}[c]{@{}c@{}}Target \\ Layout\end{tabular} \\ \hline\hline
\begin{tabular}[c]{@{}c@{}}PRIMAL [DAC'20] \\ APOLLO [MICRO'21]\end{tabular} &   \multirow{5}{*}{RTL} &  Yes &  Yes &  No &  \multirow{5}{*}{No} \\ \cline{1-1} \cline{3-5}
\begin{tabular}[c]{@{}c@{}}Sengupta et al. [ICCAD'22]\\SNS [ISCA'22]\\SNS V2 [MICRO'23]\end{tabular} &   &  No &  \multirow{3}{*}{No} &  \multirow{3}{*}{Yes} &   
\\ \cline{3-3}  \cline{6-6}
MasterRTL [ICCAD'23]&                      & Yes          &                     &                      &     \multirow{2}{*}{Yes}                 \\ \cline{1-3} 
PowPredicCT [DAC'24] &  Layout &  No &   &   &   \\ \hline\hline
ATLAS &  Netlist &  \textbf{Yes} &  \textbf{Yes} &  \textbf{Yes} &  \textbf{Yes} \\ \hline
\end{tabular}
}
\begin{tablenotes}\footnotesize
\item $^*$GRANNITE estimates toggle rate instead of power, thus not listed in the table. It is neither time-based nor targeting layout. 
\end{tablenotes} 
 \caption{Summary of representative ML-based power models.} 
\label{tbl:prior work}
 \vspace{-.2in}
\end{table}

\textbf{Our Solution ATLAS.} In this work, we propose ATLAS, targeting efficient evaluation of the time-based post-layout power of any given gate-level netlist. To the best of our knowledge, ATLAS is the first work that supports both \emph{time-based} power simulation and general \emph{cross-design} power modeling. Moreover, ATLAS targets the most accurate power label from \emph{layout stage} and \emph{realistic designs} (e.g., out-of-order CPU designs instead of small blocks). Compared with standard simulation flow based on EDA tools, the general ATLAS solution bypasses both the layout process and time-based power simulation, achieving significant speedups.

ATLAS achieves unprecedented general time-based power modeling based on a customized \emph{pre-training} and \emph{fine-tuning} paradigm. 
It proposes the following novel strategies: 
\begin{itemize}
    \item \textbf{Sub-module generation:} ATLAS first splits each design into non-overlapping sub-modules. ATLAS will evaluate the per-cycle post-layout power of each small sub-module.  
    \item \textbf{Pre-training:} A general netlist encoder model is pre-trained based on multiple self-supervised learning tasks without power labels. The encoder, based on graph transformer, is trained by guessing the masked toggle rate and node type, as well as recognizing the alignment between regular post-synthesis netlist\footnote{In this manuscript, if not explicitly described, by default ``netlist'' refers to the post-synthesis gate-level netlist before layout. The post-layout netlist (with timing optimization and clock tree) will be explicitly mentioned.} and the ultimate post-layout netlist. The encoder will encode each sub-module into a general embedding (i.e., vector) with rich design information. Such an informative embedding effectively supports challenging power modeling tasks. 
    \item \textbf{Fine-tuning:} Based on the embedding from the pre-trained encoder, we fine-tune different lightweight models for three power groups: combinational logic, register, and clock tree. 
\end{itemize}

We evaluate ATLAS on different designs with 300K to 600K cells under realistic workloads. 
ATLAS achieves high accuracy in per-cycle power modeling, with only $<1\%$ error percentage on average. ATLAS is up to $1000\times$ faster than the traditional commercial flow by bypassing both the layout process and standard time-based power simulation. The results indicate the superior predictive capability of ATLAS, demonstrating its effectiveness in cross-stage, cross-design, and time-based power prediction.

\section{Methodology Overview}\label{sec:method1}

This section provides an overview of ATLAS. ATLAS includes three major steps: design netlist preprocessing (Sec.~\ref{sec:step1} and Fig.~\ref{fig:method_sub}), pre-training (Sec.~\ref{sec:step2} and Fig.~\ref{fig:method_pre}), and fine-tuning (Sec.~\ref{sec:finetune} and Fig.~\ref{fig:method_fine}). 


\textbf{Preprocessing} (Sec.~\ref{sec:step1}). The flow begins with netlist preprocessing, preparing the dataset for ATLAS pre-training. In this step, we split each design in the gate-level netlist into many non-overlapping sub-modules. We transform each sub-module into the graph format and annotate related features (e.g., cell type, cell power from liberty file) to each node (i.e., cell instance). During training, the pre-processed data will be adopted to pre-train the encoder. During inference, ATLAS will encode each sub-module and predict its per-cycle power value.



\textbf{Pre-training} (Sec.~\ref{sec:step2}). 
The pre-training step will train a general circuit encoder, targeting two general learning goals: 1) recognizing the structure and functionality of netlists; 2) learning the alignment between netlist and layout. This encoder is pre-trained with five novel and customized self-supervised tasks to recognize the inherent semantics and structures of sub-modules, without relying on any power labels. After the pre-training, the encoder is expected to capture the transformations made on each sub-module (e.g., buffer insertion, reconstruction, clock tree) during layout and encode the information into its output embedding vector.

Unlike previous self-supervised learning tasks on netlists~\cite{Deepgate1,deepgate2,functionDAC2022} that only target combinational circuits, we target realistic sequential designs. Moreover, we incorporate per-cycle toggle information into one of the five self-supervised learning tasks, allowing ATLAS to learn about signal propagation.

\textbf{Fine-tuning} (Sec.~\ref{sec:finetune}). The final step of ATLAS fine-tunes different lightweight models for three different power groups: combinational logic, sequential logic, and clock network. These models leverage the embedding vectors generated during pre-training as input, generating fine-grained per-cycle power value of each power group for each sub-module. Summing up all power groups of all sub-modules provides the total power for each cycle.

\section{Netlist Preprocessing}
\label{sec:step1}

The netlist preprocessing step is essential for preparing the dataset used in ATLAS. This involves generating netlists from different stages and collecting key features that facilitate effective learning.

\subsection{Sub-module Format}
\label{sec:seg}

\begin{figure}[!tb]
\centering
\includegraphics[width=0.48\textwidth]{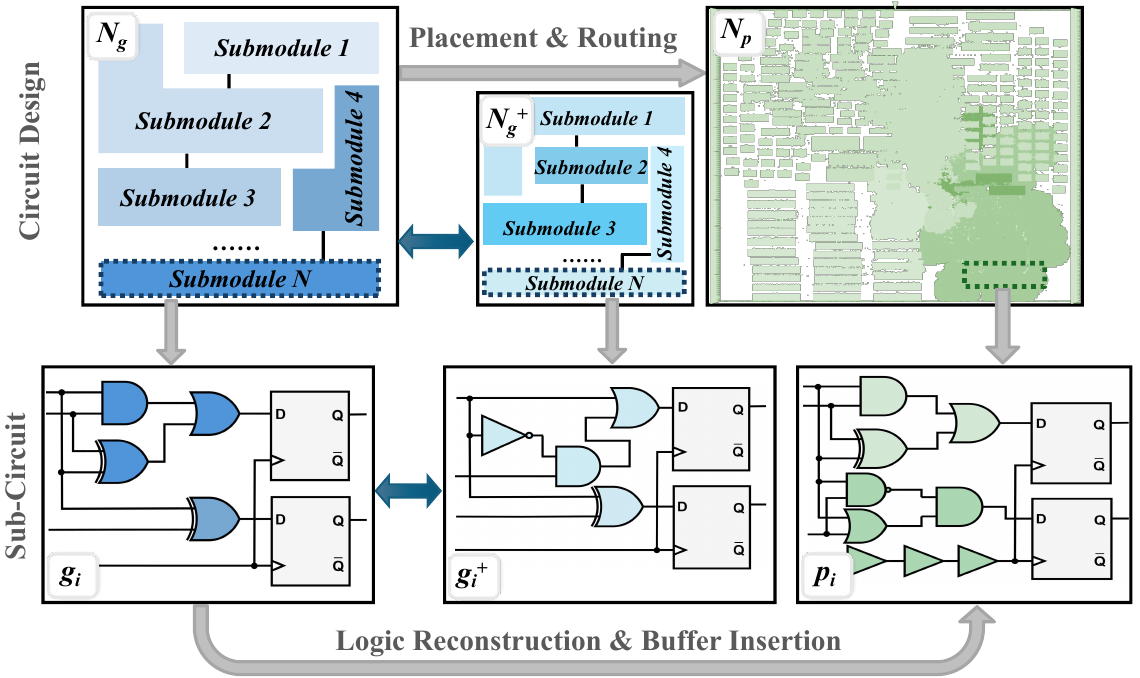}
\caption{Netlist Preprocessing of ATLAS. The preprocess includes circuit segmentation in both gate-level netlist ${N_g}$, ${N_{g}^+}$ and post-layout netlist ${N_p}$. ${N_{g}^+}$ represents a Boolean-equivalent expression of ${N_g}$ and both are functionally equivalent with similar circuit structures. For each sub-circuit ${g_i}$ in ${N_g}$, there exists an equivalent gate-level expression ${g_{i}^+}$ and a transformed expression ${p_i}$ after layout.\vspace{-.1in}}
\vspace{-.1in}
\label{fig:method_sub}
\end{figure}

Fig.~\ref{fig:method_sub} illustrates the netlist preprocessing process. Given a gate-level netlist ${N_g}$, we split it into a set of non-overlapping sub-modules ${N_g\in\{g_1,g_2,...,g_i\}}$. Then this original netlist ${N_g}$ will be transformed to a corresponding post-layout netlist ${N_p}$, which will similarly yield a set of sub-modules ${N_p\in\{p_1,p_2,...,p_i\}}$. During this standard layout process, the netlist will be optimized for better timing through buffer insertion, netlist reconstruction, etc. Also, the clock tree will be synthesized and added to the design.


Most previous works on circuit quality prediction~\cite{fang2023masterrtl, xu2022sns, xu2023fast} choose to split circuits into multiple \emph{logic cones}, then model the PPA of each logic cone. Each input code corresponds to one flip-flop, including all input logic and flip-flops that drive it. In comparison, we split the circuit into sub-modules, and then encode and evaluate the power of each sub-module. This approach offers obvious advantages:




\textbf{Non-overlapping:} 
Works~\cite{fang2023masterrtl, xu2022sns, xu2023fast} based on logic cones unavoidably involve significant overlapping between different cones, making them actually inappropriate for power modeling. 
The significant logic overlap will make the summing up of the power of each cone much larger than the total design power. 
In contrast, our approach uses non-overlapping sub-modules defined by specific functional roles. 
We can accurately estimate the power of a larger component or the whole design by summing the powers of all its constituent sub-modules.

\textbf{Fine-grained Analysis:} 
By splitting based on sub-module, the ATLAS model achieves fine-grained predictions not only in terms of time but also at the component level. Since each sub-module is determined by the inherent functionality of the netlist, the power of each sub-module provides fine-grained feedback to designers.






\subsection{Netlist Collection for Self-supervised Learning} 
\label{sec:Netlist}

To support the learning of the general intrinsic netlist information across stages, we will collect different netlists during preprocessing. 
The collected netlists will be applied in self-supervised encoder pre-training in the next step. As shown in Fig.~\ref{fig:method_sub}, we will generate three types of netlist for model pre-training: ${N_{g}}$, ${N_{p}}$ and ${N_{g}^+}$. The ${N_{g}}$ is the original gate-level netlist by synthesizing RTL code. The ${N_{g}^+}$ and ${N_{p}}$ are introduced below.


\textbf{1) For recognition of structure and functionality:} 
By applying logic-invariant transformations on ${N_{g}}$, we can derive another gate-level netlist ${N_{g}^+}$, which has the same functionality but a different structure. 
As a basic property of circuits, different gate-level netlists can implement the same functionalities.
Our encoder model will learn the {structural and functional similarities} among sub-modules by recognizing structures with the same functionality. During encoder pre-training, as we will introduce, ${N_{g}}$ provides original samples, while ${N_{g}^+}$ provides corresponding positive samples. Sub-module that differs from the target sub-module serves as a negative sample.


\textbf{2) For alignment with the layout stage:} 
We employed commercial tools to perform layout on ${N_{g}}$, with each layout step involving optimization. Ultimately, we obtained the post-routing design layout, and we refer to the corresponding post-layout netlist as ${N_{p}}$. During encoder pre-training, as we will introduce, the objective is to learn the cross-stage alignment between ${N_{g}}$ and ${N_{p}}$, making the embedding of the gate-level netlist ${E_{g}}$ to capture and reflect the information of post-layout netlist ${E_{p}}$.

\subsection{Feature Collection}
\label{sec:gate type}


We split all netlists from Section~\ref{sec:Netlist} into sub-modules.
Then we convert each sub-module to a directed graph (DG). Each {node} corresponds to each cell instance and directed edges are the wires connecting these nodes.
By representing sub-modules as DGs, we will leverage the advantages of graph transformer models to capture the circuit structures and semantics.


After converting all sub-modules to DGs, we proceed to collect features for every graph node. To benefit the subsequent learning process, we have carefully selected four types of features: {Node Type}, {Per-cycle Toggle}, {Cell Internal} and {Leakage Power}. 

\textbf{1) Node Type:} We categorize all cells according to their functionality into 18 types, using one-hot encoding to represent the cell type. 
ATLAS will learn to recognize the node type and node type effectively conveys each cell's functional characteristics. 


For example, the clock tree in post-layout netlist $N_p$ involves the interconnection of many clock-related cells, such as \emph{Clock Buffer}, \emph{Clock AND}, and \emph{Clock Multiplexer}, all categorized as the \texttt{CK} type by us. By learning to recognize masked \texttt{CK}-type nodes during pre-training, the encoder is guided to capture the effect of the clock tree.


\textbf{2) Per-cycle Toggle:}
Per-cycle toggle represents switching activity at a node's output port during a specific timestamp under real workloads.
Including a per-cycle toggle aims for ATLAS to learn the signal propagation relation between nodes and to predict signal propagation. During the pre-training, the encoder will learn by predicting the masked toggle behavior.

\textbf{3) Cell Internal and Leakage Power:}
For each cell instance, both internal and leakage power values will be parsed from the lookup tables in the \texttt{.lib} files from the technology library according to its cell type. These standard cell power values provide important cell information and are directly power-related. 


\section{ATLAS Pre-training}
\label{sec:step2}


\begin{figure}[!tb]
\centering
\includegraphics[width=0.5\textwidth]{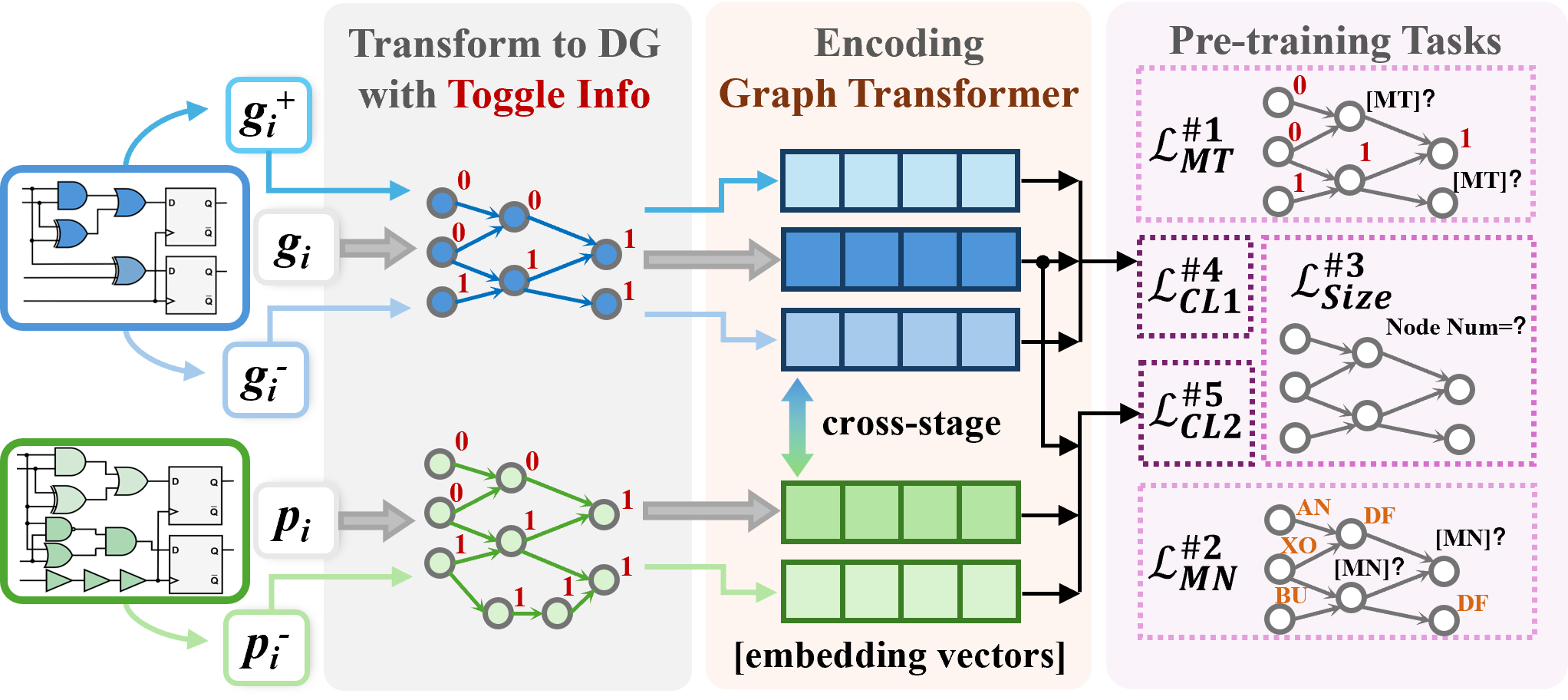}
\caption{ATLAS Pre-training. This framework employs a self-supervised and cross-stage flow that incorporates two types of contrastive learning. Additionally, ATLAS includes predictions for [$\mathtt{MASK\_TOGGLE}$] ([MT]) sub-module size and [$\mathtt{MASK\_NODE\_TYPE}$] ([MN]). \vspace{-.1in}}
\vspace{-.1in}
\label{fig:method_pre}
\end{figure}

ATLAS trains a general circuit encoder to convert each sub-module into a representation.  
Fig.~\ref{fig:method_pre} illustrates the entire self-supervised pre-training process of the encoder model based on efficient graph transformer~\cite{wu2023sgformer}. The encoder will generate an embedding for each individual node, as well as an additional overall graph embedding for the whole input graph (i.e., corresponding sub-module).



The encoder is pre-trained on the unlabeled circuit dataset, targeting the following two general learning goals: 
1) recognizing the structure, functionality, and toggle propagation of netlists; 2) learning the alignment between netlist and layout. 
The encoder is expected to encode the learned information into its output embedding vector.
To achieve this, we carefully designed five encoder pre-training tasks: \ding{182} Masked toggle propagation learning; \ding{183} Masked node type learning; \ding{184} Sub-module size learning; \ding{185} Gate-level netlist contrastive learning; \ding{186} Cross-stage alignment contrastive learning.

Our proposed five self-supervised learning tasks can be roughly categorized into three types, as summarized below. 
\begin{enumerate}
    \item Masked node recovery (Tasks \ding{182}, \ding{183}). We will mask (i.e., hide) important node properties (e.g., type, toggle) of randomly selected nodes. A temporary MLP-based classification head will predict the masked node properties based on the encoder-generated node embedding. During pre-training, the encoder and classification head (to be discarded after pre-training) will be trained end-to-end to maximize accuracy. The encoder will thus learn to encode the structural information of connected circuit operators within the circuit graph. 
    \item Size recognition (Task \ding{184}). A temporary MLP-based regression head will predict the total number of nodes in the given graph (i.e., sub-module netlist) based on encoder-generated graph embedding. During pre-training, the encoder and regression head (to be discarded after pre-training) will be trained end-to-end to maximize accuracy. The encoder will learn to encode more global graph information in the overall graph embedding. 
    \item Netlist contrastive learning (Tasks \ding{185}, \ding{186}). The encoder will be trained to minimize the embedding distance of netlists of the same sub-module with the same functionality (i.e., positive samples), while maximizing the embedding distance of different sub-modules (i.e., negative samples).
\end{enumerate}

We formally introduce and formulate each learning task below.

\textbf{Task \#1 Masked toggle propagation learning:}  To train our encoder model to handle per-cycle workloads, we mask the per-cycle toggle information of randomly selected nodes. A temporary MLP classifier will predict whether the masked node is toggling based on encoder-generated node embeddings. 
This pre-training task trains the encoder to capture the toggle information and the propagation of toggles. Specifically, we randomly mask the toggle feature (0 or 1) of selected nodes and replace it with a special token [$\mathtt{MASK\_TOGGLE}$]. Subsequently, we predict whether these nodes will toggle based on the embedding of the masked node, which also captures neighboring nodes' information. We represent the input circuit in masked graph format as $\hat{G}$. The ground-truth toggle feature of the masked nodes is represented by ${\boldsymbol{t}}^{msk}_{\hat{G}}$. The predicted toggle feature for the masked nodes is represented by ${\boldsymbol{p}}^{msk}({\hat{G}})$, which is based on node embeddings\footnote{To simplify the formulation, the encoder model does not directly appear in the loss term in Eq.~(\ref{loss1}). The prediction ${\boldsymbol{p}}^{msk}({\hat{G}})$ is based on encoder model.}. The objective is to minimize the ross-entropy (CE) loss between ${\boldsymbol{t}}^{msk}_{\hat{G}}$ and ${\boldsymbol{p}}^{msk}({\hat{G}})$, as expressed in the following formula:
\begin{equation}
\mathcal{L}^{\#1}_{\text{MT}} = \mathbb{E}_{\hat{G} \sim \mathcal{D}} CE  \left( {\boldsymbol{t}}^{msk}_{\hat{G}}, \  {\boldsymbol{p}}^{msk}({\hat{G}}) \right)  \label{loss1}
\end{equation}
where $\mathbb{E}_{\hat{G} \sim \mathcal{D}}$ represents the expectation $\mathbb{E}$ over the circuit graph dataset $\mathcal{D}$.

\textbf{Task \#2 Masked node type learning:}
Similar to the toggle task, we randomly mask the node type information and replace it with a special token [$\mathtt{MASK\_NODE\_TYPE}$]. A temporary MLP classifier will predict the one-hot encoding for the masked node types based on the embeddings of the masked node. The input circuit's masked representation is denoted as \( \hat{G} \), with the ground-truth type for the masked nodes represented by \( {\boldsymbol{c}}^{\text{msk}}_{\hat{G}} \) and the predicted type denoted by \( {\boldsymbol{q}}^{\text{msk}}(\hat{G}) \). The objective is to minimize the cross-entropy (CE) loss between the true one-hot encoding \( {\boldsymbol{c}}^{\text{msk}}_{\hat{G}} \) and the predicted one-hot encoding \( {\boldsymbol{q}}^{\text{msk}}(\hat{G}) \), expressed as:
\begin{equation}
\mathcal{L}^{\#2}_{\text{MN}} = \mathbb{E}_{(E_g) \sim \mathcal{D}} {CE}\left( {\boldsymbol{c}}^{\text{msk}}_{\hat{G}}, \ {\boldsymbol{q}}^{\text{msk}}(\hat{G})) \right)
\end{equation}


\textbf{Task \#3 Sub-module size learning:}
The size of a circuit design is often directly correlated with its overall power consumption. 
We propose a pre-training task to recognize the size of each sub-module. A temporary MLP regressor will predict the number of nodes ($\text{Num}_{pre}$) in the whole graph (i.e., cell count in sub-module) based on the overall graph embedding. The objective is to minimize the MSE between predicted node number $\text{Num}_{pre}$ and the actual number of nodes $\text{Num}_{true}$:
\begin{equation}
\mathcal{L}^{\#3}_{\text{Size}} = \mathbb{E}_{\hat{G} \sim \mathcal{D}} \left[ \left( \text{Num}_{pre} - \text{Num}_{true} \right)^{2} \right]
\end{equation}

\textbf{Task \#4 Gate-level netlist contrastive learning:}
To capture the functionalities of each sub-module, we employed contrastive learning on the netlist. Specifically, for the gate-level netlist ${N_{g}}$ with many sub-modules $g_i$, we generate another ${N_{g}^+}$ through functionally equivalent transformations. ${N_{g}^+}$ provides new sub-modules ${g_{i}^+}$ with identical functionalities but different structures. At this task, we consider ${g_{i}}$ and ${g_{i}^+}$ as a pair of positive samples. All other sub-modules in the batch that exhibit different functionalities are treated as negative samples (${g_{i}^-}$) of ${g_{i}}$. The contrastive objective pulls functionally similar sub-modules closer together in their respective embedding spaces while pushing dissimilar sub-modules further apart. We denote the embeddings of ${g_{i}}$, ${g_{i}^+}$, and ${g_{i}^-}$ as ${E_{g}}$, ${E_{g}^+}$, and ${E_{g}^-}$, respectively. Specifically, we minimize the InfoNCE loss~\cite{infonce} used for sub-module embeddings, defined as $\text{CL}$, as follows:
\begin{equation}
\mathcal{L}^{\#4}_{\text{CL1}} = \mathbb{E}_{(E_g) \sim \mathcal{D}} {CL}\left( {E_{g}}, {E_{g}^+} ,{E_{g}^-}\right)
\end{equation}

\textbf{Task \#5 Cross-stage alignment contrastive learning:}
For cross-stage contrastive learning, similar to Task \#4, we aim for gate-level netlist ${N_{g}}$ to be closer in the embedding space to the corresponding post-layout netlist ${N_{p}}$. This alignment task targets capturing layout information from ${N_{p}}$ in netlist embedding. For such contrastive learning, the original sample is still ${g_{i}}$, while the positive sample is ${p_{i}}$, and the negative samples ${p_{i}^-}$ comprise all other functionally distinct sub-modules in the batch. We denote the embeddings of ${p_{i}}$ and ${p_{i}^-}$ as ${E_{p}}$ and ${E_{p}^-}$, respectively. We also adopt the InfoNCE loss function for this purpose:
\begin{equation}
\mathcal{L}^{\#5}_{\text{CL2}} = \mathbb{E}_{(E_{g},E_p) \sim \mathcal{D}} {CL}\left( {E_{g}},{E_{p}},{E_{p}^-}\right)
\end{equation}

Finally, we formulate the complete self-supervised pre-training objective of our model by jointly employing the five tasks: 
\begin{equation}
\mathcal{L} = \mathcal{L}^{\#1}_{\text{MT}} + \mathcal{L}^{\#2}_{\text{MN}} + \mathcal{L}^{\#3}_{\text{Size}} + \mathcal{L}^{\#4}_{\text{CL1}} + \mathcal{L}^{\#5}_{\text{CL2}} 
\end{equation}

Regarding the encoder of ATLAS, traditional Graph Neural Networks (GNNs)~\cite{xu2018how} and Graph Transformers~\cite{wu2023difformer}~\cite{NodeFormer} can be resource-intensive when applied to large-scale graphs. Considering that some of our DGs contain tens of thousands of nodes, we employ an efficient Graph Transformer, SGFormer~\cite{wu2023sgformer}, as the encoder for ATLAS. In our DGs, SGFormer achieves propagation capabilities across any node using a single global attention network of linear complexity, while maintaining linear complexity with respect to the number of nodes. Furthermore, it does not require positional encodings, feature or graph preprocessing, or additional loss functions.


\vspace{-.1in}
\section{ATLAS Fine-tuning}
\label{sec:finetune}
\begin{figure}[!tb]
\centering
\includegraphics[width=0.48\textwidth]{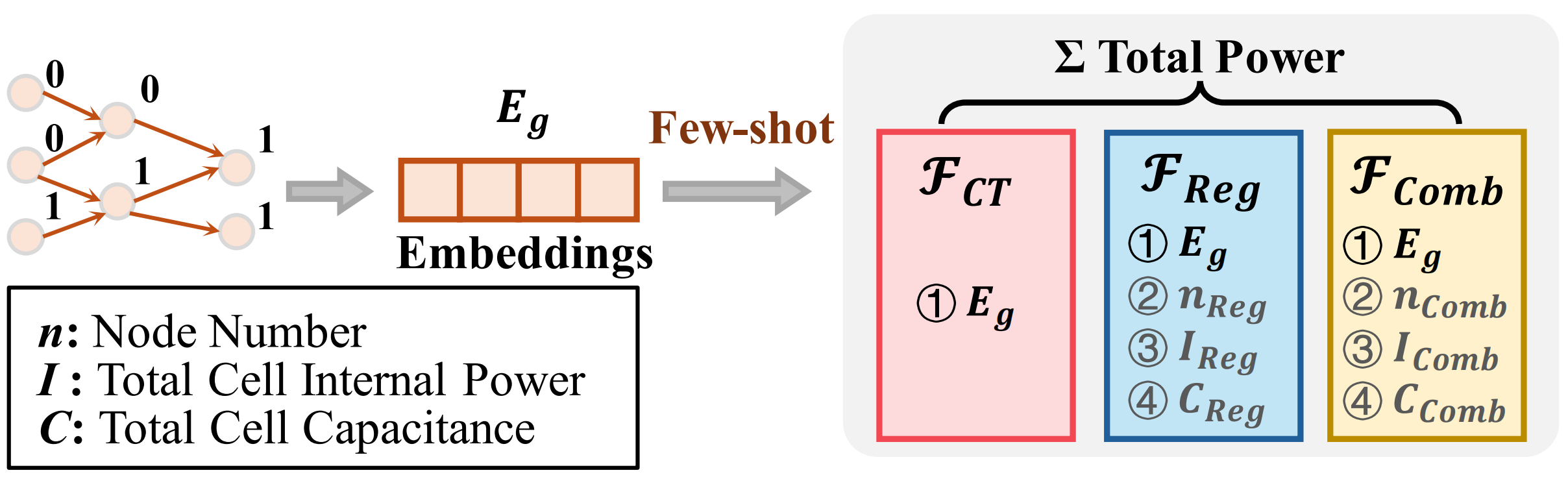}
\caption{ATLAS Fine-Tuning. The total post-layout power is divided into different groups as three separate fine-tuning tasks, which are predicted by three fine-tuning models. Other than embedding vectors ${E_g}$, ATLAS uses cell internal power, cell capacitance, and node number as features, avoiding the need for cumbersome feature engineering and complex machine learning models.\vspace{-.1in}}
\vspace{-.1in}
\label{fig:method_fine}
\end{figure}

To apply pre-trained ATLAS on downstream power modeling tasks, as shown in Fig.~\ref{fig:method_fine}, we divide the total power into three power groups: the {clock tree} power, the {combinational logic} power, and the {register} power\footnote{Our register power group includes (and is dominated by) the power of the clock pin in each register. Accordingly, our clock tree power group does not include such register clock pin power.}. Each power group will be modeled by its respective fine-tuning model, denoted as $\mathcal{F}_{CT}$, $\mathcal{F}_{Comb}$, and $\mathcal{F}_{Reg}$. 
These fine-tuned power models differ from purely supervised methods by utilizing the embedding ${E_g}$ from the encoder as features, rather than relying on cumbersome feature engineering and complex ML models to process raw circuits.

\begin{figure*}[!t]
\vspace{-.15in}
\centering
\includegraphics[width=1\textwidth]{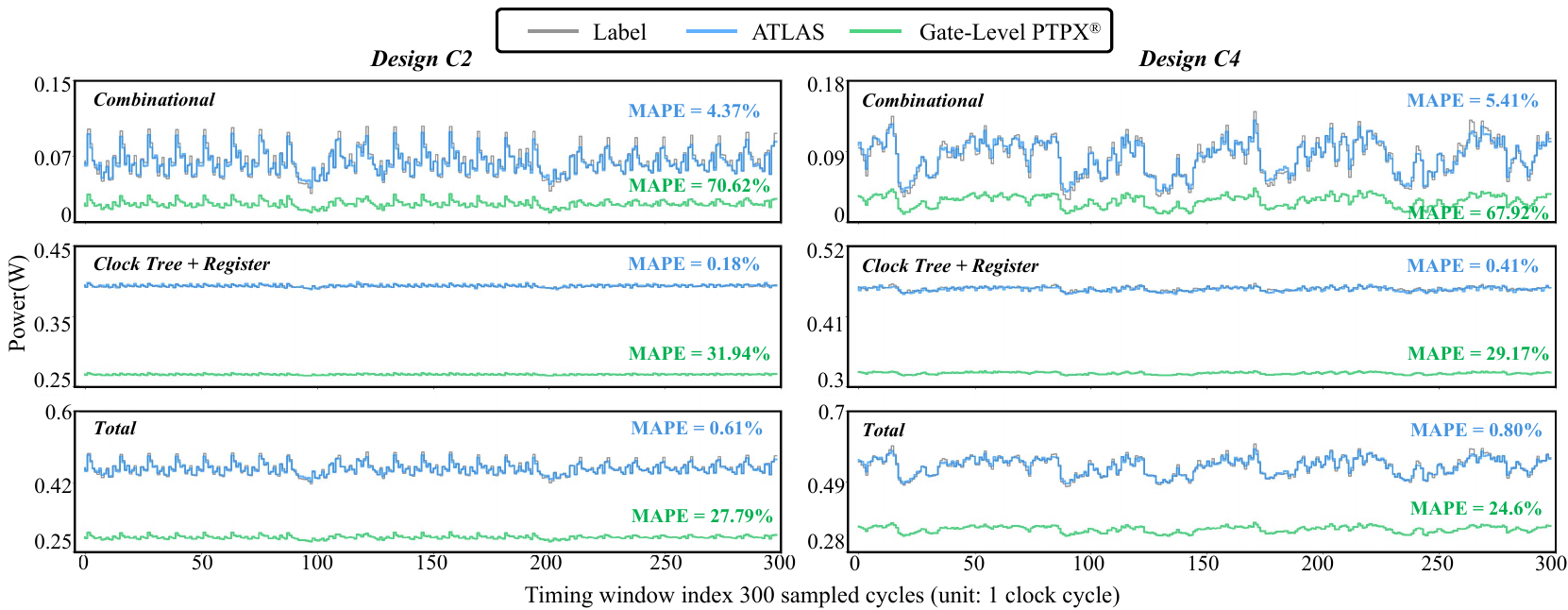}
\caption{Power prediction results from ATLAS across 300 cycles for \textit{C2} and \textit{C4} under $W_1$. The labels were obtained by simulating the netlist and SPEF files using PTPX\textsuperscript{\textregistered} after detailed routing with Innovus\textsuperscript{\textregistered}. The Gate-Level PTPX\textsuperscript{\textregistered} values, which are obtained from the gate-level netlist, illustrate the significant differences in netlist power between gate level and layout.\vspace{-.1in}}
\vspace{-.15in}
\label{fig:Power1}
\end{figure*}

Regarding the clock tree group, it is completely absent in the post-synthesis netlist ${N_g}$ and appears only in post-layout netlist ${N_p}$, which is unknown during inference. Therefore, we do not incorporate any additional features in addition to the embedding. We rely solely on the embedding ${E_g}$ to predict the power of the clock tree group. This approach will demonstrate that our encoder model has effectively learned the alignment between ${N_g}$ and ${N_p}$.



For both combinational and register groups, we choose to combine the embeddings $E_g$ with additional relevant features available from netlist $N_g$. 
For the combinational group, we utilize the number of combinational nodes $n_{Comb}$ of the post-synthesis netlist, along with the total cell internal power ${I_{Comb}}$ and total cell capacitance ${C_{Comb}}$ of the combinational group as features. The cell internal power and cell capacitance of each node are collected from the lookup tables in the \texttt{.lib} file of the technology library. To calculate the total cell internal power ${I_{Comb}}$ and total cell capacitance ${C_{Comb}}$, we multiply each cell internal power and capacitance in the combinational section by the per-cycle toggle of their output pins and sum up the results.



As for the register group, similar to the combinational group, we adopt the number of register nodes $n_{Reg}$, the total cell internal power ${I}_{Reg}$ and total cell capacitance ${C}_{Reg}$ of the register group from $N_g$ as additional features for fine-tuning. 

 As a result, our fine-tuning models are lightweight and easily integrable, including tree-based models, such as XGBoost~\cite{chen2016xgboost}. 
 So the total predicted power is:
\begin{equation}
\begin{aligned}
P_{total} &= \mathcal{F}_{CT}({E_g}) + \mathcal{F}_{Comb}({E_g}, n_{Comb}, {I_{Comb}}, {C_{Comb}}) \\
          &\quad + \mathcal{F}_{Reg}({E_g}, n_{Reg}, {I_{Reg}}, {C_{Reg}})
\end{aligned}
\end{equation}

\section{Experimental Results} \label{sec:expr}

\subsection{ATLAS Implementation and Experiment Data Generation}\label{section5-1}

We implement our ATLAS using PyTorch and PyTorch Geometric (PyG), a library built on top of PyTorch that facilitates writing and training GNNs. Our pre-training is conducted on a Linux machine equipped with an NVIDIA A6000 GPU, while the fine-tuning tasks are performed on an Intel Xeon processor with 128GB of memory.

RTL simulation on workloads is based on Synopsys VCS\textsuperscript{\textregistered}\cite{vcs} (VCS). The logic synthesis is executed at a clock frequency of 1GHz using Synopsys Design Compiler\textsuperscript{\textregistered}\cite{design-compilier} (DC). In our experiments, we utilized the TSMC 40nm standard cell library~\cite{URL:tmsc40nm}, along with the corresponding Memory Compiler.
We utilized the Innovus\textsuperscript{\textregistered} to perform mixed-size placement, clock tree synthesis, and routing, with each step including timing optimization, ultimately resulting in the chip design layout. After detailed routing with Innovus\textsuperscript{\textregistered}, we dump out the post-layout ultimate netlist and corresponding RC values in Standard Parasitic Extraction Format (SPEF) files. We conduct accurate layout-stage power simulation based on the post-layout netlist, SPEF file, and different workloads (denoted as $W_1$, $W_2$). The per-cycle ground-truth layout power simulation is performed based on Synopsys PrimeTime PX\textsuperscript{\textregistered}~\cite{ptpx} (PTPX\textsuperscript{\textregistered}).


\begin{table}[!b]
\renewcommand{\arraystretch}{1.1}
\vspace{-.15in}
 \resizebox{.485\textwidth}{!}{
 \begin{tabular}{|c||c|c|c|c|c|c|}
\hline
                    & \textit{\textbf{C1}} & \textit{\textbf{C2}} & \textit{\textbf{C3}} & \textit{\textbf{C4}} & \textit{\textbf{C5}} & \textit{\textbf{C6}} \\ \hline \hline
\textbf{Gate-level} & 289384               & 322664               & 389120               & 399486               & 465129               & 597877               \\ \hline
\textbf{Post-layout}  & 301650               & 340923               & 412186               & 422391               & 494614               & 638666               \\ \hline
\end{tabular}}
\caption{The statistics of the gate counts for the six designs at the gate-level and post-layout stages.}
\label{tbl:size}
\vspace{-.10in}
\end{table}

In this work, we generated a dataset with six different realistic designs (named C1 to C6 from smallest to largest) that support workload simulation. Table \ref{tbl:size} presents the cell count statistics for six different designs at both the gate-level and post-layout stages. The cell counts range from approximately 300,000 to 600,000, indicating their difference. For the same design, the increase in cell count from the gate-level stage to the post-layout stage reflects the impact of timing optimization and clock tree synthesis during the layout process.


\begin{table*}[!tb]
\vspace{-.2in}
\renewcommand{\arraystretch}{1.3}
    \centering
    \resizebox{.99\textwidth}{!}{
    \begin{tabular}{|cc|ccccc||ccccc|}
\hline
\multicolumn{2}{|c|}{\multirow{2}{*}{\textbf{Design \& Workload}}} &
  \multicolumn{5}{c||}{\textbf{Error Percentage of ATLAS}} &
  \multicolumn{5}{c|}{\textbf{Error Percentage of Gate-Level PTPX\textsuperscript{\textregistered}}} \\ \cline{3-12} 
\multicolumn{2}{|c|}{} &
  \multicolumn{1}{c|}{\textbf{Combinational}} &
  \multicolumn{1}{c|}{\textbf{Clock Tree}} &
  \multicolumn{1}{c|}{\textbf{Register}} &
  \multicolumn{1}{c|}{\textbf{Clock Tree + Register}} &
  \textbf{Total} &
  \multicolumn{1}{c|}{\textbf{Combinational}} &
  \multicolumn{1}{c|}{\textbf{Clock Tree}} &
  \multicolumn{1}{c|}{\textbf{Register}} &
  \multicolumn{1}{c|}{\textbf{Clock Tree + Register}} &
  \textbf{Total} \\ \hline \hline
\multicolumn{1}{|c|}{\multirow{2}{*}{\textit{\textbf{C2}}}} &
  \textit{\textbf{$W_1$}} &
  \multicolumn{1}{c|}{4.37\%} &
  \multicolumn{1}{c|}{0.17\%} &
  \multicolumn{1}{c|}{0.27\%} &
  \multicolumn{1}{c|}{0.18\%} &
  0.61\% &
  \multicolumn{1}{c|}{70.62\%} &
  \multicolumn{1}{c|}{100\%} &
  \multicolumn{1}{c|}{2.36\%} &
  \multicolumn{1}{c|}{31.94\%} & 
  27.79\%
   \\ \cline{2-12} 
\multicolumn{1}{|c|}{} &
  \textit{\textbf{$W_2$}} &
  \multicolumn{1}{c|}{5.35\%} &
  \multicolumn{1}{c|}{0.15\%} &
  \multicolumn{1}{c|}{0.32\%} &
  \multicolumn{1}{c|}{0.22\%} &
  0.57\% &
  \multicolumn{1}{c|}{71.01\%} &
  \multicolumn{1}{c|}{100\%} &
  \multicolumn{1}{c|}{2.22\%} &
  \multicolumn{1}{c|}{31.93\%} & 27.81\%
   \\ \hline
\multicolumn{1}{|c|}{\multirow{2}{*}{\textit{\textbf{C4}}}} &
  \textit{\textbf{$W_1$}} &
  \multicolumn{1}{c|}{5.41\%} &
  \multicolumn{1}{c|}{1.07\%} &
  \multicolumn{1}{c|}{0.54\%} &
  \multicolumn{1}{c|}{0.41\%} &
  0.80\% &
  \multicolumn{1}{c|}{67.92\%} &
  \multicolumn{1}{c|}{100\%} &
  \multicolumn{1}{c|}{2.34\%} &
  \multicolumn{1}{c|}{29.17\%} & 24.60\%
   \\ \cline{2-12} 
\multicolumn{1}{|c|}{} &
  \textit{\textbf{$W_2$}} &
  \multicolumn{1}{c|}{5.34\%} &
  \multicolumn{1}{c|}{0.93\%} &
  \multicolumn{1}{c|}{0.68\%} &
  \multicolumn{1}{c|}{0.67\%} &
  1.13\% &
  \multicolumn{1}{c|}{69.36\%} &
  \multicolumn{1}{c|}{100\%} &
  \multicolumn{1}{c|}{2.44\%} &
  \multicolumn{1}{c|}{29.23\%} & 25.08\%
   \\ \hline \hline
\multicolumn{2}{|c|}{\textbf{Average}} &
  \multicolumn{1}{c|}{5.12\%} &
  \multicolumn{1}{c|}{0.58\%} &
  \multicolumn{1}{c|}{0.45\%} &
  \multicolumn{1}{c|}{0.37\%} &
  0.78\% &
  \multicolumn{1}{c|}{69.73\%} &
  \multicolumn{1}{c|}{100\%} &
  \multicolumn{1}{c|}{2.34\%} &
  \multicolumn{1}{c|}{30.57\%} & 26.32\%
   \\ \hline
    \end{tabular}
     }
    \caption{MAPE (\%) of design \textit{C2} and \textit{C4} under two workloads $W_1$ and $W_2$.\vspace{-.18in}}
    \label{tbl:vvadd}
\end{table*}


For ATLAS pre-training stage, our dataset comprises aligned 3,253 DG pairs of ${N_{g}}$, ${N_{g}^+}$, ${N_{g}^-}$ and ${N_{p}}$ for each cycle. 
We employed four designs (\textit{C1}, \textit{C3}, \textit{C5} and \textit{C6}) for training and two designs (\textit{C2} and \textit{C4}) for testing, ensuring absolute no overlap between the training and testing designs. Ultimately, the number of DGs in the training and testing datasets is 92,500 and 37,640, respectively. During the encoder pre-training process, the five self-supervised tasks are trained simultaneously for 60 epochs, requiring approximately 21 hours. Despite such training runtime, please notice that the encoder only needs to be pre-trained once and then applied to different designs. We employ ReLU as the activation function and set the batch size to 16, with a learning rate of 1e-4 using the Adam optimizer. In the fine-tuning phase, we use XGBoost with 500 estimators and a depth of 5, taking only several seconds for training.

\subsection{Power Modeling Experiment Setup}

\emph{Commericial Tool as Baseline.} Our experiment will evaluate the modeling accuracy and efficiency of per-cycle post-layout power based on post-synthesis netlists (without any layout information). Moreover, the tested design is strictly not \emph{seen} by the power model during training. As summarized in the introduction, prior works either require a design-specific model for time-based power modeling~\cite{zhou2019primal, xie2021apollo} or can only evaluate the average power~\cite{xu2022sns, fang2023masterrtl, xu2023fast, sengupta2022good, du2024powpredi}. As a result, no prior ML-based power models apply to this challenging cross-design per-cycle power modeling task. Moreover, it is very challenging to naturally adapt prior works to our brand-new tasks not supported originally. Therefore, we choose to compare ATLAS with the standard commercial tool at the post-synthesis netlist stage.


\emph{Exclusion of Memory Group from Results.} In our analysis of power consumption, memory (i.e., SRAM) accounts for almost half of the total power in the entire design. But for this memory power group, we can predict its power with very high accuracy, without too much effort at the netlist stage. Since the SRAM macro is unchanged during layout, we developed a basic ML model based on the toggle activities of memory ports from workload, and energy values from lookup tables in SRAM \texttt{.lib} files. Our model achieves an error of only 0.5\% without relying on any power simulators. Given such high accuracy in memory power prediction, incorporating the memory power model into ATLAS would lead to a lower error, but primarily dominated by the memory group. To demonstrate the benefits of our ATLAS effectively, we choose to exclude this easier memory group in results. Instead, we focus on predicting the more challenging other powers of the combinational, clock tree, and register groups.
As introduced Section~\ref{sec:finetune}, we implemented three corresponding fine-tuned models: $\mathcal{F}_{CT}$, $\mathcal{F}_{Comb}$, and $\mathcal{F}_{Reg}$.

We evaluate the accuracy of per-cycle power modeling with Mean Absolute Percentage Error (MAPE) between label $Y_i$ and prediction $\hat{Y}_i$, assuming altogether $m$ cycles in the tested workload.  
\begin{equation}
\text{MAPE} = \frac{1}{m} \sum_{i=1}^{m} \left| \frac{Y_i - \hat{Y}_i}{Y_i} \right| \times 100\%
\end{equation}

\vspace{-.18in}
\subsection{Power Prediction Results}
\vspace{-.03in}

Fig.~\ref{fig:Power1} visualizes power prediction results (ATLAS) and the actual power values (Labels) over 300 cycles for testing designs \textit{C2} and \textit{C4}, under workload $W_1$. 
For comparison, we present the commercial tool's simulation result on exactly the same gate-level netlist and workload (Gate-Level PTPX\textsuperscript{\textregistered}). 
For both \textit{C2} and \textit{C4}, the overall power trace shapes of ATLAS across the three power groups closely resemble the labels, with the total MAPE being only 0.61\% and 0.80\%, respectively. In comparison, the gate-level PTPX\textsuperscript{\textregistered} error is higher than 25\% for both designs. Its power trace also differs significantly from the label. This error shows the gap between the power of the gate-level netlist and the power of the ultimate layout.

 

In Table~\ref{tbl:vvadd}, we provide a more comprehensive comparison of results for ATLAS and Gate-Level PTPX\textsuperscript{\textregistered}, which displays the MAPE for different power groups under both workloads $W_1$ and $W_2$.
Gate-Level PTPX\textsuperscript{\textregistered} still exhibits the $>25\%$ prediction accuracy in total power in all cases, while ATLAS shows $<1\%$ error. 

In terms of the most challenging combinational group prediction, ATLAS shows an average MAPE of 5\%, while Gate-Level PTPX\textsuperscript{\textregistered} has a considerably higher MAPE of 69\%. Since there is no clock tree in the gate-level netlist, Gate-Level PTPX\textsuperscript{\textregistered} shows a MAPE of 100\%. ATLAS demonstrates a remarkably low MAPE of only 0.58\%.

In summary, standard commercial tools at the gate-level stage show $>25\%$ error on time-based layout power due to the lack of layout information. ATLAS, by pre-training and fine-tuning, well captures the potential impact of layout and demonstrates a remarkable $<1\%$ error for the per-cycle power of new circuit design.

\begin{figure}[!t]
\centering
\includegraphics[width=0.48\textwidth]{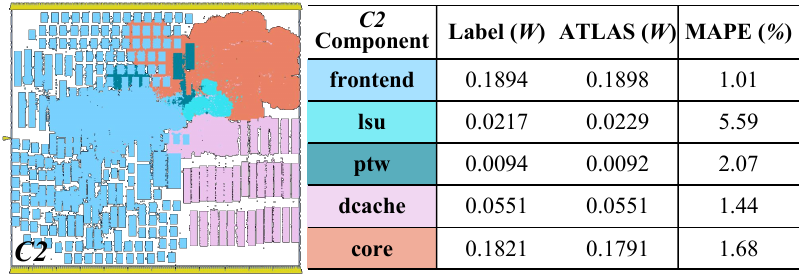}
\caption{Component-level power analysis of \textit{C2} under $W_1$, illustrating the division into five major components, each comprised of their respective sub-modules, highlighting ATLAS's accuracy in the component level. C2 is an out-of-order CPU design, its main components include the CPU frontend, load-store unit (LSU), data cache (dcache), etc.}
\vspace{-.05in}
\label{fig:result2}
\end{figure}

Fig.~\ref{fig:result2} further illustrates the ATLAS-predicted power of five major components in the design $C2$. The prediction of each component's power is the summation of the predicted power of all its sub-modules. Combined with the layout, the component power predictions further provide power distributions on the layout. 
In the table on the right of Fig.~\ref{fig:result2}, we present each component's labels, ATLAS power predictions, and the associated MAPE values.
It is evident that the power predictions vary across different components, with some over-estimation and others under-estimation. The component error is slightly higher than total power but mostly maintained $<5\%$. 

\subsection{Runtime Comparison}
\vspace{-.03in}
Table~\ref{tbl:time} presents the time taken for back-end processing using Innovus, the time for power simulations using PTPX\textsuperscript{\textregistered}, and the time taken by ATLAS to directly predict power. Here we included the runtime used for the entire preprocessing step in ATLAS. We calculated the average time (in seconds) for six designs. Compared with the complete traditional flow that performs both complete layout and time-based power simulations, ATLAS finished estimating the 300-cycle workload with $>1000\times$ less runtime.


In summary, compared to standard design flow based on commercial tools, ATLAS offers three significant advantages: \ding{182} it eliminates the need for the time-consuming layout process; \ding{183} it reduces the time required for per-cycle power simulation; \ding{184} the accuracy of power predictions for post-layout netlists by ATLAS far exceeds that of Gate-Level PTPX\textsuperscript{\textregistered} at the early-stage of gate-level netlists.

%


\begin{table}[]
\renewcommand{\arraystretch}{1.2}
\centering
\resizebox{.45\textwidth}{!}{
\begin{tabular}{|c|ccc|ccc|}
\hline
\multirow{2}{*}{\textbf{Design}} & \multicolumn{3}{c|}{\textbf{ATLAS}}                                                       & \multicolumn{3}{c|}{\textbf{Traditional Flow}}                                          \\ \cline{2-7} 
                                 & \multicolumn{1}{c|}{\textbf{Pre.}} & \multicolumn{1}{c|}{\textbf{Infer}} & \textbf{Total} & \multicolumn{1}{c|}{\textbf{P\&R}} & \multicolumn{1}{c|}{\textbf{Simulation}} & \textbf{Total} \\ \hline\hline
\textit{C1}   & \multicolumn{1}{c|}{62}& \multicolumn{1}{c|}{3.2}&     65.2           & \multicolumn{1}{c|}{46392}         & \multicolumn{1}{c|}{96}&   46488\\ \hline
\textit{C2}   & \multicolumn{1}{c|}{66}& \multicolumn{1}{c|}{4.8}&       70.8         & \multicolumn{1}{c|}{55327}         & \multicolumn{1}{c|}{98}&   55425\\ \hline
\textit{C3}    & \multicolumn{1}{c|}{67}& \multicolumn{1}{c|}{4.1}&        71.1        & \multicolumn{1}{c|}{69624}         & \multicolumn{1}{c|}{105}&  69709\\ \hline
\textit{C4}    & \multicolumn{1}{c|}{72}& \multicolumn{1}{c|}{4.6}&     76.6           & \multicolumn{1}{c|}{79836}         & \multicolumn{1}{c|}{112}&  79948\\ \hline
\textit{C5}    & \multicolumn{1}{c|}{75}& \multicolumn{1}{c|}{5.1}&        80.1       & \multicolumn{1}{c|}{105742}        & \multicolumn{1}{c|}{131}&  105873\\ \hline
\textit{C6}     & \multicolumn{1}{c|}{86}& \multicolumn{1}{c|}{4.9}&      90.9         & \multicolumn{1}{c|}{124860}        & \multicolumn{1}{c|}{156}&  125016\\ \hline\hline
Average     & \multicolumn{1}{c|}{72}              & \multicolumn{1}{c|}{4}               &    \textbf{76}            & \multicolumn{1}{c|}{80297}& \multicolumn{1}{c|}{116}&     \textbf{80413}\\ \hline
\end{tabular}
}
\caption{Runtime (in seconds) comparison for 300 cycles extracted from $W_1$. Column names represent: Pre. (Data Preprocessing), Infer (Inference), Simulation (Time-based Power Simulation).\vspace{-.2in}} 
    \label{tbl:mape vvadd}
\label{tbl:time}
\end{table}

\section{Conclusion}\label{sec:concl}

In this paper, we propose ATLAS, a pioneering framework for fine-grained per-cycle power prediction. ATLAS is the first work that supports both time-based power simulation and general cross-design power modeling. It achieves unprecedented general time-based power modeling based on a customized pre-training and fine-tuning paradigm. When evaluated on the prediction of per-cycle post-layout power based on gate-level netlist, ATLAS achieves a remarkable MAPE of $<1\%$ for total power estimation with inference speeds that are $>1000\times$ faster than traditional flow.


\section{Acknowledgments}\label{sec:acknow}

This work is supported by National Natural Science Foundation of China (NSFC) 62304192, Hong Kong Research Grants Council (RGC) ECS Grant 26208723, and ACCESS–AI Chip Center for Emerging Smart Systems, sponsored by InnoHK, Hong Kong SAR. 
We thank HKUST Fok Ying Tung Research Institute and National Supercomputing Center in Guangzhou Nansha Sub-center for computational resources.

\bibliographystyle{IEEEtran}
\bibliography{ref}

\end{document}